# Energy Levels and Transition Rates of Laser-Cooling Candidate Th⁻

Rui Zhang, Yuzhu Lu, Shuaiting Yan, and Chuangang Ning[a)]

*Department of Physics, State Key Laboratory of Low Dimensional Quantum Physics, Frontier Science Center for Quantum Information, Tsinghua University, Beijing 100084, China*

[a)] Author to whom correspondence should be addressed: ningcg@tsinghua.edu.cn

**ABSTRACT**

The energy levels and transition rates of promising laser-cooling candidate thorium anion Th⁻ have been experimentally obtained in the current work. Three new excited states between the bound-bound electric dipole (E1) transition $^2S^o_{1/2} \leftrightarrow {}^4F^e_{3/2}$ of Th⁻ are observed now: $^4G^o_{5/2}$, $^4F^e_{9/2}$ and $^4F^o_{7/2}$, and their energy levels are determined to be 1,847(13), 3,166.8(59) and 3,666(12) cm⁻¹, respectively. Meanwhile, the lifetime of the upper state $^2S^o_{1/2}$ of that E1 transition is experimentally determined to be 30(2) μs, about 3 times shorter than the previous calculated result 86 μs, which makes Th⁻ the most promising candidate for laser cooling of negative ions. Furthermore, the lifetimes of two other short-lived odd-parity excited states of Th⁻ are also measured to be 59(4) and 53(3) μs, respectively.

## I. INTRODUCTION

Laser cooling [1,2], a crucial experimental technique for producing ultracold atoms and molecules, is opening up lots of brand-new research areas [3,4] and enabling experiments at unprecedentedly low temperatures [5,6]. Initially achieved in positive ions Ba⁺ [7] and Mg⁺ [8] in 1978, the technique was later extended to neutral atom Na [9]. However, this promising technology has remained confined to neutral and positive ions systems, with no realization in any negative ions over the past half-century. The primary obstacle is the lack of a closed and fast electric-dipole (E1) cycling transition between two bound states within the anionic candidate systems for the laser cooling. In contrast to positive ions and neutral systems, which typically have nearly infinite bound states, negative ions possess their extra electrons by the short-range polarization or electron correlation effects [10]. As a result, anions are generally characterized by having no excited bound states, possessing only one ground state [11-14]. Negative ions with excited bound states are very rare, and the presence of E1 cycling transitions between these states is even more uncommon. The molecular anion $C_2^-$ has been proposed as a candidate for laser cooling [15-17]. However, molecular cooling requires the recycling of vibrational and rotational branching during the cycling transition, complicating the laser systems involved.

Only four atomic anions (Os$^-$ [18-23], Ce$^-$ [24,25], La$^-$ [22,26-29] and Th$^-$ [30,31]) have been discovered to possess the rare E1 transitions found in the entire periodic table. Among them, Ce$^-$ has numerous metastable lower states known as "dark states" between its bound-bound transition, making it less suitable for anionic laser cooling. The other three candidates have their own strengths and weaknesses. Os$^-$ was the first anion with an observed bound E1 transition [18]. The wavelength of its cooling transition has been measured at 1,162.75 nm in the near-infrared range [20], which is comparatively more accessible than the mid-infrared ranges of La$^-$ (3,103.7 nm) [27,29] and Th$^-$ (2,428.4 nm) [31]. However, our recent experiments determined the lifetime of the upper excited state of its E1 transition to be 201(10) μs [23], an order of magnitude slower than the theoretical rates of the other two candidate anions La$^-$ and Th$^-$. La$^-$ was calculated to has the faster E1 transition rate at $4.9 \times 10^{-4}$ s$^{-1}$ and the shortest cooling time of 1.2 s among all the candidates [29]. However, its stable isotope $^{139}$La$^-$ with a nuclear spin of 7/2 has nine hyperfine components between its bound-bound cycling transition, requiring a complicated laser system for repumping [29].

For Th$^-$, its potential laser-cooling transition cycle $^2S_{1/2}^o \leftrightarrow {}^4F_{3/2}^e$ and a few anionic bound states were predicted by the Multiconfiguration Dirac-Hartree-Fock (MCDHF) method in 2019 [30]. On the experimental side, the electron affinity of Th has been determined to be 4,901.35(48) cm$^{-1}$ using the cryo-SEVI spectrometer [30]. However, only partially predicted bound states ($^4F_{3,5,7/2}^e$) have been experimentally observed by the direct photodetachment [30]. In 2021, three electric-dipole transitions (T1, T2 and T3) from its ground stated of Th$^-$ to the short-lived odd-parity states have been observed through a resonance scan [31]. T1 corresponds to the transition to the upper state $^2S_{1/2}^o$ of laser-cooling transition cycle of Th$^-$, with an energy level 4,118.0(10) cm$^{-1}$ above its ground state. T2 and T3 have been tentatively assigned from odd-parity states $^4F_{5/2}^o$ and $^4D_{1/2}^o$, with energies of 4,592.6(10) and 4,618.1(10) cm$^{-1}$, respectively. The complex electronic structure of Th$^-$ remains incompletely understood, and several predicted states within the E1 transition range have yet to be observed. Further investigation is required to check their impact on the optical cycling of Th$^-$.

Currently, the lifetime of the upper state $^2S^o_{1/2}$ of this E1 cooling transition is based solely on a theoretical estimate of 86 μs [31] and a transition rate of $1.17 \times 10^{-4}$ s$^{-1}$. The absence of direct experimental measurements underscores the need for the ongoing research. More detailed comparative data on these three negative-ion laser-cooling candidates can be found in the Supplemental Table I.

In the present work, we report the observation of three newly bound excited states of Th$^-$ ($^4G^o_{5/2}$, $^4F^e_{9/2}$ and $^4F^o_{7/2}$) by extending the range of our detachment laser into the infrared band. The relatively long lifetime of the $^2S^o_{1/2}$ state of Th$^-$ (~ a few tens of microseconds) renders the typical pump-probe method unfeasible. Utilizing a cold ion trap, the first pulsed laser excites Th$^-$ in the ion trap, and after an adjustable delay, a second pulsed laser measure the population in the excited using the photoelectron energy spectroscopy. The lifetimes of three short-lived odd-parity excited states of Th$^-$ were successfully measured.

## II. EXPERIMENTAL METHODS

In brief, our cryo-SEVI apparatus has four main parts: an ion source for producing ions, a cryogenic ion trap for trapping anions, a time-of-flight (TOF) mass spectrometer for selecting Th$^-$, and a velocity-map imaging system for photodetachment [31-34]. A pulsed 532-nm Nd: YAG laser was focused onto a thorium metal target for ablation to produce ions in the source chamber. The resulting negative ions were guided by a hexapole and trapped in the radio frequency (RF) octupole cryogenic ion trap [35], and where they were cooled down to a nominal temperature of a few K via the collisions with the buffer gas (H$_2$ or He) delivered by a pulsed valve. The typical trapping period is ~ 45 ms. A mid-infrared laser at 2,428.4-nm (4,118.0 cm$^{-1}$) from a difference frequency generation (DFG) system was used to excite Th$^-$ from the ground $^4F^e_{3/2}$ state to the $^2S^o_{1/2}$ state in the cryogenic ion trap through a CaF$_2$ view window. The DFG system covers an infrared tuning range of 1.5 − 4.2 μm, achieved with a dye laser (Spectra-Physics, linewidth ~0.06 cm$^{-1}$) and its residual fundamental 1064-nm output. The wavelength and intensity were monitored by a wavelength meter (HighFinesse WS6-600) with an accuracy of 0.02 cm$^{-1}$. After an adjustable delay, the excited Th$^-$ ions were ejected out from the ion trap for population analysis using the photoelectron energy spectroscopy. Given that the time of flight in the TOF mass spectrometer [36] is a fixed value for Th⁻, the adjustable "*delay time*" (the time interval between the excitation laser fired on and the ion trap ejection) was used to measure the lifetime of the excited states of Th⁻ (See Figs.S1 in supplementary information). To minimize the de-excitation of the excited Th⁻ during the TOF flight, the acceleration voltage of TOF was increased to 3 kV from 1 kV, resulting in a reduction of the TOF time to 33 µs from 57 µs. The ejection voltage of the ion trap was also increased to reduce the time of excited Th⁻ to reach the accelerating electrodes. In the velocity-imaging system [37,38], an optical parametrical oscillator (OPO) laser (Spectra-Physics primoScan, 400 – 2,700 nm, linewidth ~ 6 cm⁻¹) and the dye laser (400 – 920 nm) were used for the photodetachment to measure the energy levels of Th⁻. The apparatus operated at a repetition rate of 20 Hz.

The distribution of detached photoelectron was reconstructed using the maximum entropy velocity Legendre reconstruction (MEVELER) method [39,40]. Since the detached electrons with the same velocity could form a spherical shell with the same radius ($r$), its kinetic energy ($E_k$) can be given by $E_k = \alpha r^2$. $\alpha$ is the calibration coefficient of the spectrometer. The radius $r$ was determined through Gaussian function fitting of the center position of each peak. The energy levels obtained at different photon energies were further optimized using a global optimization analysis [41,42].

### III. RESULTS AND DISCUSSION

**Three new bound states**

Figure 1 shows photoelectron spectra of Th⁻ obtained through direct detachment at different photodetachment wavelengths. The energy spectrum in black was acquired at photon energy 11,603.57 cm⁻¹ using the dye laser, while the others were recorded using the DFG laser. The strongest peak *9* is the transition from the anionic ground state to the neutral ground state, and its binding energy was measured to be 4,902.6(61) cm⁻¹, consistent with the previous result of the electron affinity (EA) of Th atom [30]. At lower photon energies, additional weaker peaks were resolved. Peaks *4* and *5* have similar photoelectron kinetic energies but show distinctly different photoelectron angular distributions, suggesting that their initial anionic states possess opposite parities. Since peak *5* has been identified as the transitions from the anionic excited state Th⁻ ($^4F^e_{5/2}$) of to the atomic ground state Th ($^3F_2$) [30], peak *4* is assigned to the transition from the lowest anionic odd-parity state Th⁻ ($^4G^o_{5/2}$) to the same neutral state Th ($^3F_2$) now. This transition was predicted by high-precision calculations [30] and had not been experimentally observed previously due to the resolution limitation. To assign these peaks from anionic excited states, we changed the trap time to analyze the intensity trends of the different peaks. The intensity of a peak originating from an excited state with a shorter lifetime exhibit faster intensity decay with increasing trap time. We collected a series of energy spectra with different trapped time (blue curve for 45ms, purple for 5ms) and the no-trap mode (red for no-trap). Peaks *1* and *6* correspond to be the transition from the new anionic state $^4F^o_{7/2}$ to the Th ground state. Peak *3* was previously assigned as the transition from the long-lived anionic excited state Th⁻ ($^4F^e_{7/2}$) [30]. And peaks *2*, *7* and *10* are now assigned to the transitions from the same new anionic state $^4F^e_{9/2}$ to different neutral Th states. Note that the peak *8* near peak EA in Fig.2 is a transition from Th⁻ ($^4F^e_{7/2}$) to Th ($^3F_3$), with a measured binding energy of 4878.8(61) cm⁻¹. The energy gap between peak *8* and *9* is only 20 cm⁻¹, highlighting the high resolution near the photodetachment threshold. Since the binding energy of this transition *8* has been precisely measured near the threshold, the energy level of the anionic state $^4F^e_{7/2}$ is also updated to a more accurate value to be 2892.9(50) cm⁻¹ through global optimization analysis. The binding energies and assigned of all observed peaks are listed in the Supplemental Table II. The three newly identified anionic excited states are illustrated in Fig.3 and listed in Table I.

**Transition rate of the laser cooling cycle**

To measure the transition rate of the laser cooling transition $^2S^o_{1/2} \leftrightarrow {}^4F^e_{3/2}$, we measured the lifetime of the upper state $^2S^o_{1/2}$ via the E1 transition. As shown in Fig. 2 a), no peak from $^2S^o_{1/2}$ can be observed through the single-photon direct photodetachment due to its lower population and a predicted lifetime of a few tens of microseconds [31]. To observed this short-lived state, a DFG laser with a photon energy 4,118.0 cm⁻¹ was used to excite Th⁻ to the state $^2S^o_{1/2}$. Three emerging peaks labeled *T1-1*, *T1-2* and *T1-3* were successfully observed, as illustrated by the red curve in Fig.2. These new peaks are from the excited state $^2S^o_{1/2}$ to the different final neutral states. All assigned transitions are listed in the Supplemental Table II. The relative intensity of these peaks from $^2S^o_{1/2}$ decreased gradually when increasing the delay time due to the spontaneous emission. To compensate the fluctuation of the ion beam intensity, we recorded the ratio of peak *T1-1* to strong peak *9* versus the "*delay time*" (defined in the experimental methods). The lifetime of $^2S^o_{1/2}$ state was determined to be 30(2) μs via exponential decay fitting to the observed results, as shown in Fig.2 b). This experimental result is about 3 times shorter than the previous calculation 86 μs [31], making Th⁻ as the most promising candidate for the laser cooling of negative ions. The mean free time of the buffer gas ($H_2$) collisions in the ion trap at the end of trapping was estimated to be 5 ms, much longer than the measured lifetime. To evaluate the impact of collisions on the lifetime, we reduced the gas density in the ion trap by monitoring the background pressure of the chamber from $2 \times 10^{-4}$ Pa to $1 \times 10^{-4}$ Pa, the lifetime remained at 30(2) μs (See Figs.S2 in supplementary information), indicating that collision effects are negligible. In addition, we measured the lifetimes of other two short-lived odd-parity states of Th⁻ via the resonant peaks T2 and T3, determining their lifetime to be 59(4) and 53(3) μs, respectively, as shown in Fig.4, which align closely with the previous predictions [31]. The summary of the lifetimes data of these three odd-parity states is presented in Table II.

**Discussion on laser-cooling candidate Th⁻**

With the energy levels and lifetimes of the laser-cooling candidate Th⁻ further experimentally determined, we updated the calculations related to its laser cooling. For the three newly observed excited states between the cooling cycle transition $^2S^o_{1/2} \leftrightarrow {}^4F^e_{3/2}$: $^4G^o_{5/2}$, $^4F^e_{9/2}$ and $^4F^o_{7/2}$, the transitions from the state $^2S^o_{1/2}$ to these states are E2, M4 and M3 types, respectively, which should be extremely weak [30], not breaking the closure of the laser cooling cycle. With the updated lifetime of the upper state $^2S^o_{1/2}$ now at 30(2) μs, the transition rate is calculated to be $3.3(2) \times 10^{-4}$ s$^{-1}$, approximately three times faster than previous predictions [31]. Consequently, the laser-cooling period $T$ is shortened from 7.8 s to 2.7 s, while the natural linewidth $\Delta\nu$ of this transition increases from 1.9 kHz to 5.4 kHz. The photodetachment loss rate, $\propto \Delta\nu T$, remains negligible at 0.025% [31]. The only stable isotope $^{232}$Th, with zero nuclear spin, has no hyperfine splitting in its cooling transition, greatly simplifying the cooling laser system compared to $^{139}$La⁻. A current challenge for Th⁻ is the difficulty in obtaining mid-infrared lasers (2,428.4 nm). We plan to use external-cavity diode lasers to address this issue. In principle, once Th⁻ anions are effectively cooled, the collisional sympathetic cooling [19,43] could be used to cool down any other negatively charged species. Combined with the threshold photodetachment [23], a variety of cold neutral atoms and molecules could also be produced.

## IV. CONCLUSIONS

In summary, we have experimentally measured the lifetime of the upper state for E1 transition of laser-cooling candidate thorium anion Th⁻ to be 30(2) μs. This result reveals a laser cooling rate nearly three times faster than previously predictions [31]. We also experimentally obtained the lifetimes of two other short-lived odd-parity excited states as 59(4) and 53(3) μs, respectively. The complex energy levels of Th⁻ anion have now been clarified, with the observation of three new excited states $^4G^o_{5/2}$, $^4F^e_{9/2}$ and $^4F^o_{7/2}$. Our results on energy levels and lifetimes of the thorium anions demonstrate the existence of a closed and fast E1 transition, strongly supporting Th⁻ as a promising candidate for laser cooling of negative ions. Furthermore, these experimental results also provide a benchmark for the development of high-precision theoretical calculations for heavy elements.

## ACKNOWLEDGMENTS

This work was supported by the National Natural Science Foundation of China (NSFC) (Grant Nos. 12374244 and 12341401) and the China Postdoctoral Science Foundation Grant (GZC20231367).

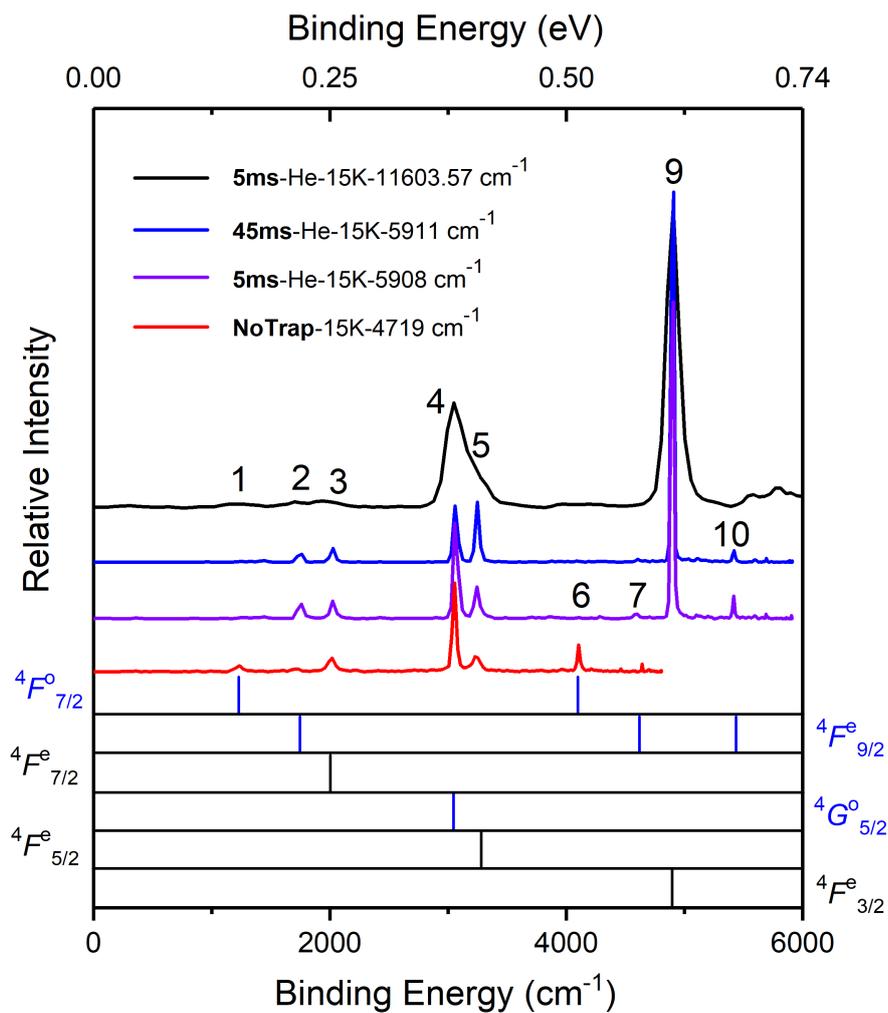

FIG. 1. The photoelectron spectra of Th⁻ observed at the different photon energies and different trapped time. The assignments of each transition labeled with numbers are listed in the Supplemental Table II. Three sets of blue sticks below the spectra indicate the energy levels of the final neutral states from the three newly identified excited states of Th⁻ with labels. The other three sets of black sticks indicate the assigned anionic states $^4F^e_{3,5,7/2}$ of Th⁻ before.

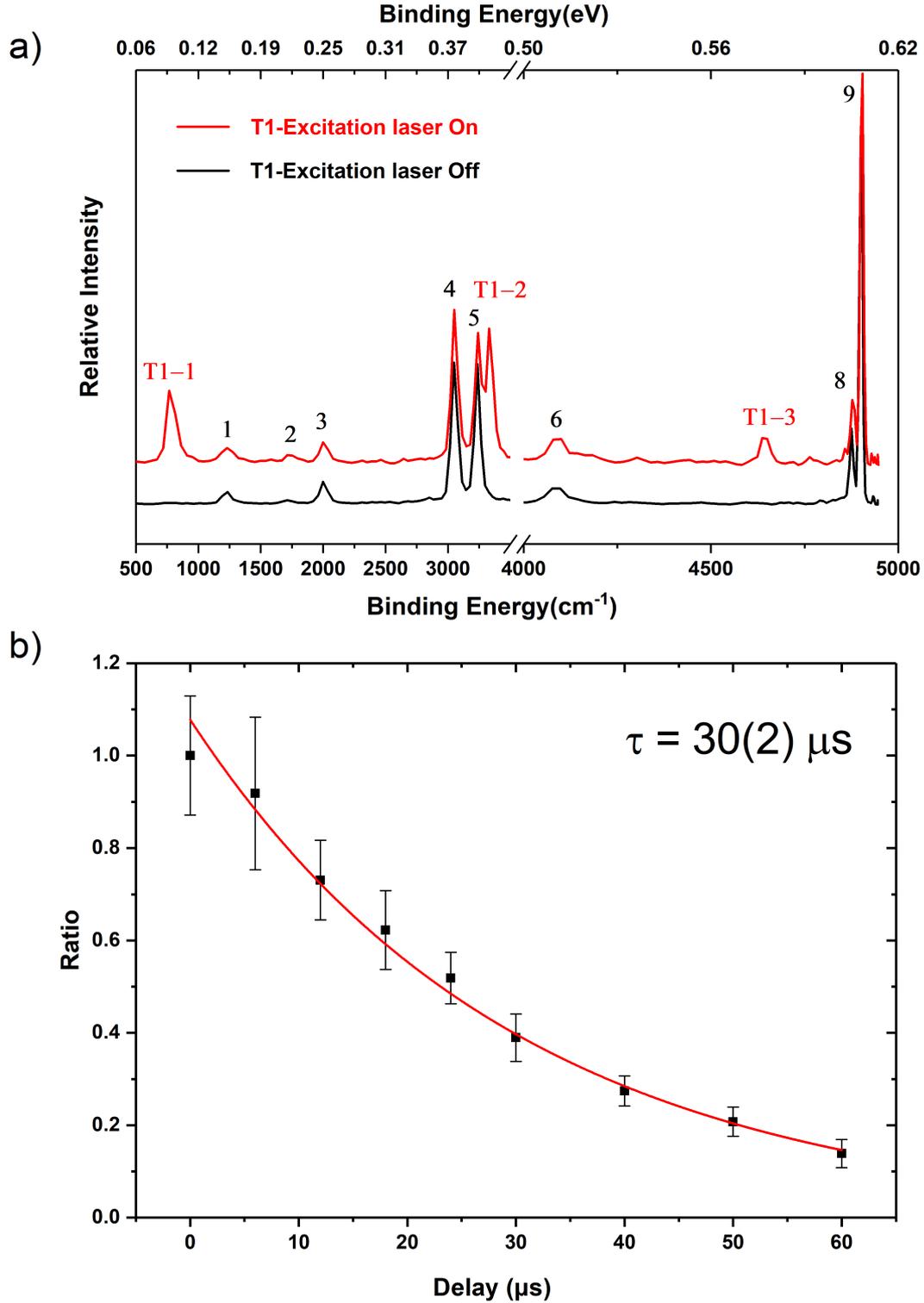

FIG. 2.  a) The photoelectron energy spectra of Th⁻ with (in red) and without (in black) the excitation laser at 4,118.0 cm⁻¹. The photon energy of the detachment laser is 4,947 cm⁻¹. The red peaks T1-1,2,3 are assigned as the transitions from the short-lived odd-parity excited state $^2S^o_{1/2}$ of Th⁻. b) The intensity ratio of transition *T1-1* Th($^3F_2$) ← Th⁻ ($^2S^o_{1/2}$) to the transition *9* Th($^3F_2$) ← Th⁻ ($^4F^e_{3/2}$) versus the delay time. The red solid line is the fitting curve according to equation $I(t) = I_0 e^{-t/\tau}$. The lifetime of $^2S^o_{1/2}$ state is measured to be 30(2) μs.

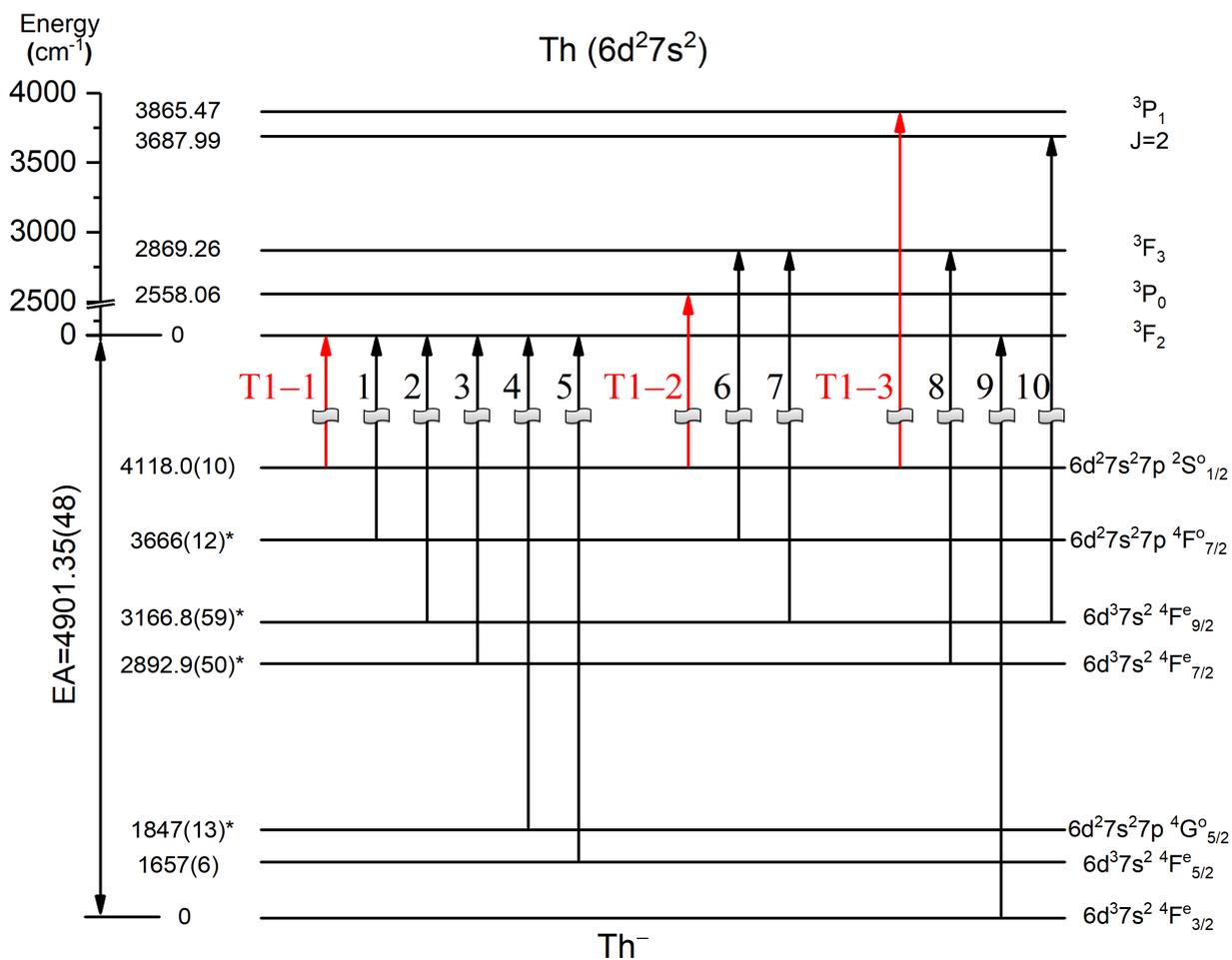

FIG. 3.  The related energy levels of Th⁻ and Th. Labels of each transition are the same as the observed peaks in Figs. 1 - 2. The data with asterisk of energy levels are determined by the present work.

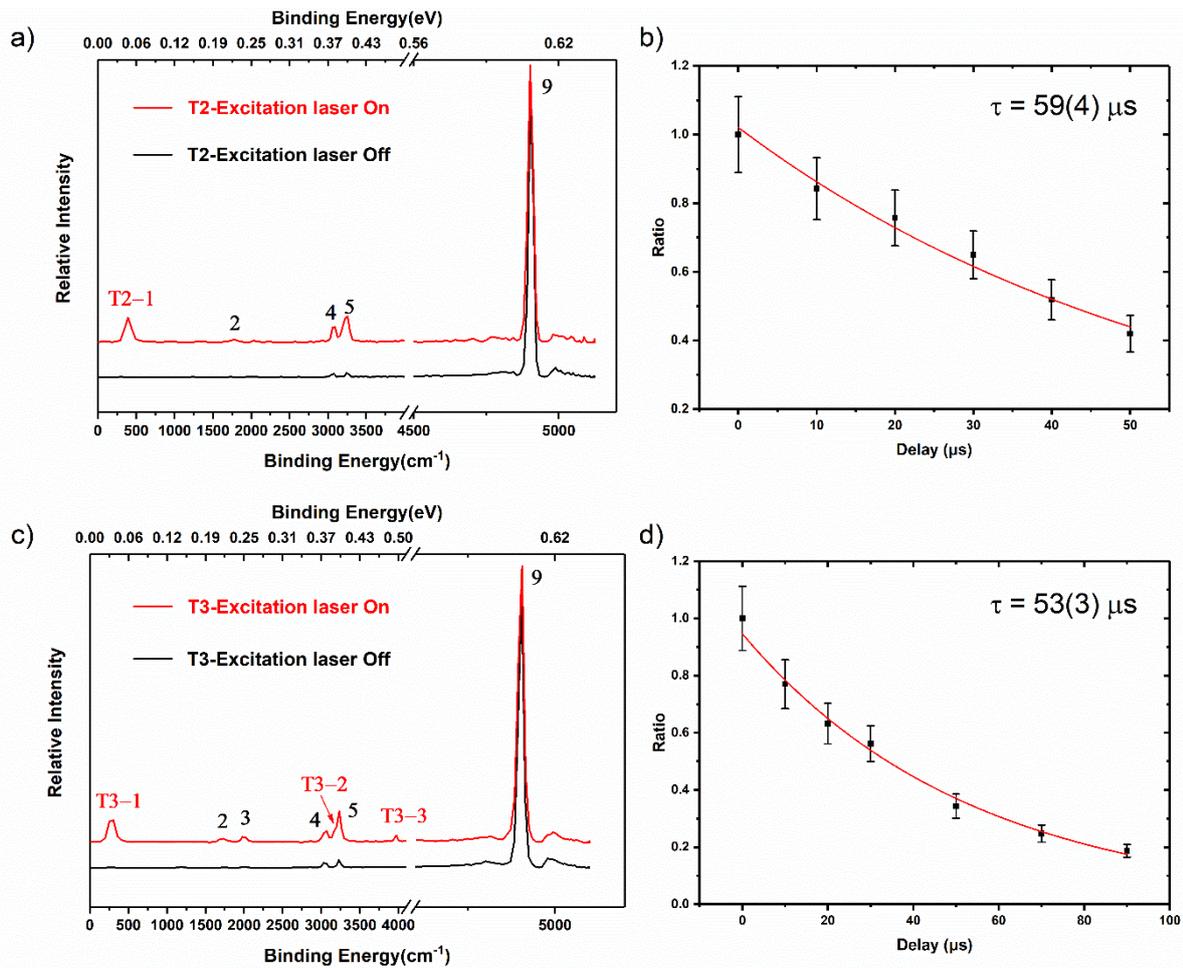

FIG. 4. The photoelectron energy spectra of Th⁻ with (in red) and without (in black) the excitation laser at 4,592.6 (a, T2) and 4,618.1 cm$^{-1}$ (c, T3). The photon energies of each detachment laser are 5,124 (T2) and 5,098 cm$^{-1}$ (T3), respectively. The intensity ratios of transition *T2-1* (b) and *T3-1* (d) to the transition *9* versus the delay time. The lifetimes of these two short-lived odd-parity states are measured to be 59(4) and 53(3) μs, respectively.

TABLE I. Summary of all the measured energy levels of Th⁻ anion.

| States of Th⁻ [a] | Energy levels (cm$^{-1}$) | References |
|---|---|---|
| $^4F^e_{3/2}$ | 0 | Tang et al. [30] |
| $^4F^e_{5/2}$ | 1657(6) | |
| $^4G^o_{5/2}$ | 1847(13) | This work |
| $^4F^e_{7/2}$ | 2892.9(50) | |
| $^4F^e_{9/2}$ | 3166.8(59) | |
| $^4F^o_{7/2}$ | 3666(12) | |
| $^2S^o_{1/2}$ | 4118.0(10) | Tang et al. [31] |
| $^4F^o_{5/2}$ | 4592.6(10) | |
| $^4D^o_{1/2}$ | 4618.1(10) | |

[a] The electronic configuration of Th⁻ are $6d^37s^2$ (even parity) and $6d^27s^27p$ (odd parity).

TABLE II. Summary of the lifetimes ($\tau$) in μs, transition energies (E) in cm$^{-1}$ and rates (A) in s$^{-1}$ of three observed short-lived odd-parity excited states of Th⁻. T1 is the laser-cooling transition cycle $^2S^o_{1/2} \leftrightarrow {}^4F^e_{3/2}$ of Th⁻. Numbers in brackets for the transition rate A represent powers of 10.

| Transitions | Type | E | A | $\tau$(Expt.) | $\tau$(Theo.) |
|---|---|---|---|---|---|
| T1($^2S^o_{1/2} \leftrightarrow {}^4F^e_{3/2}$) | E1 | 4118.0 | 3.33[4] | 30(2) | 86 |
| T2 | E1 | 4592.6 | 1.69[4] | 59(4) | 42 |
| T3 | E1 | 4618.1 | 1.89[4] | 53(3) | 44 |

# Supplementary Information

## Energy Levels and Transition Rates of Laser-Cooling Candidate Th⁻

Rui Zhang, Yuzhu Lu, Shuaiting Yan, and Chuangang Ning[a]

*Department of Physics, State Key Laboratory of Low Dimensional Quantum Physics, Frontier Science Center for Quantum Information, Tsinghua University, Beijing 100084, China*

[a] Author to whom correspondence should be addressed: ningcg@tsinghua.edu.cn

**Contents**

**Figures and Tables**

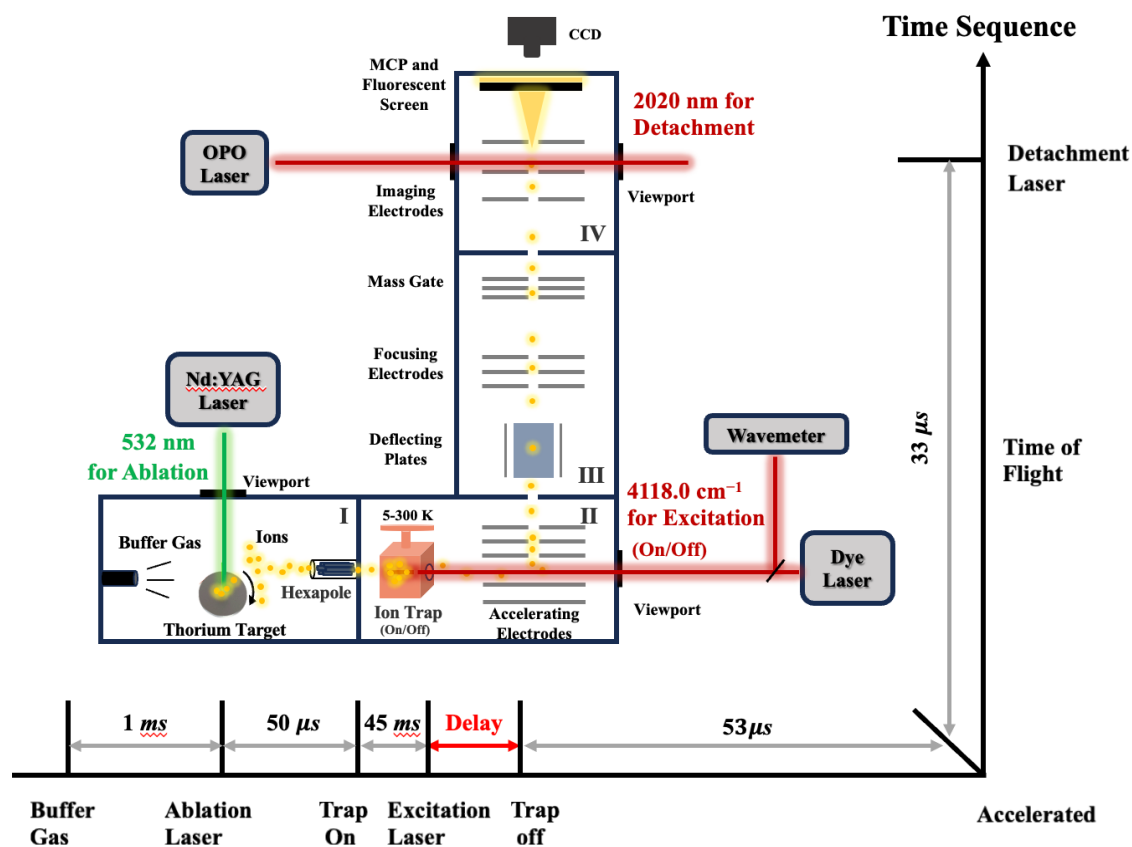

**Supplemental Figure 1.** Schematic view and the time sequence of our cryo-SEVI apparatus. Four main parts of this apparatus: I-Ion Source (Ions was generated by laser ablation), II-Cryogenic Ion Trap (Ions was trapped then excited), III-The TOF Mass Spectrometer (Thorium anion was selected) and IV-Velocity Imaging System (Laser detachment and imaging). The time sequence along with the trajectory of Th⁻ (yellow dots) outside the schematic of the apparatus. The time interval between the excitation laser on and the ion trap off called "Delay" (in red) used to measure the lifetime of the excited states of Th⁻. Details are in the article.

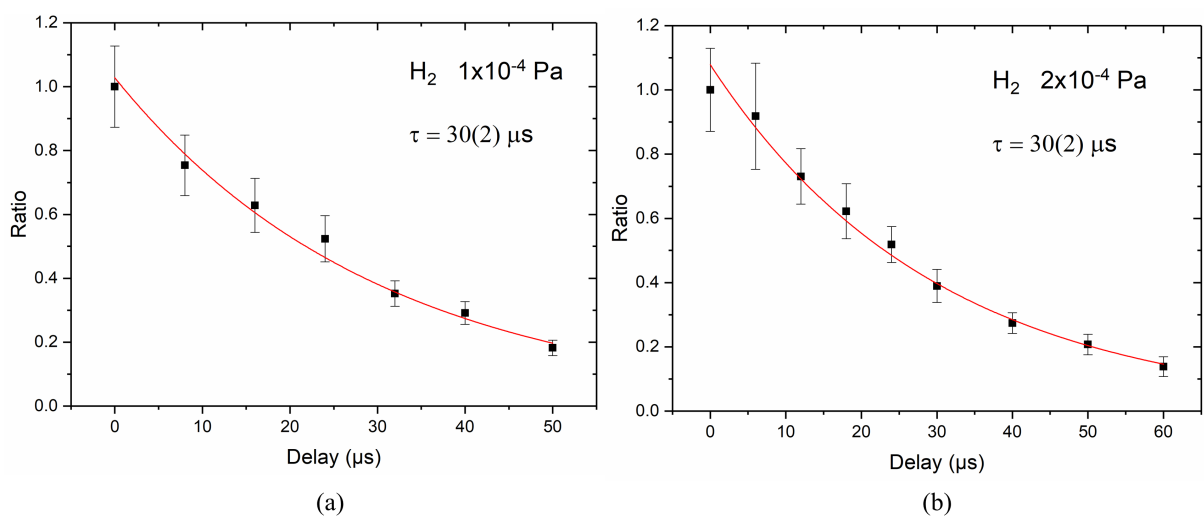

**Supplemental Figure 2.** The intensity ratio of transition Th ($^3F_2$) ← Th$^-$ ($^2S^o_{1/2}$) to transition Th ($^3F_2$) ← Th$^-$ ($^4F^e_{3/2}$) versus the delay time. The red solid line is the fitting curve by equation $I(t) = I_0 e^{-t/\tau}$. The lifetime of $^2S^o_{1/2}$ state is measured as 30(2) μs at the background pressure of both a) $1 \times 10^{-4}$ Pa and b) $2 \times 10^{-4}$ Pa.

**Supplemental Table I.** Comparative data of three anionic laser cooling candidates [1-12].

| Anions | Lifetimes (μs) | Transition Rates (s$^{-1}$) | Wavelength (nm) | Frequency (THz) | Cooling Time (s) | Repumping Lasers | Nuclear Spin |
|---|---|---|---|---|---|---|---|
| Th$^-$ | 30(2) | $3.3 \times 10^4$ | 2428.4 | 123.455 | 2.8 | 0 | 0 |
| Os$^-$ | 201(10) | $5.0 \times 10^3$ | 1162.8 | 257.831 | 4.7 | 1 | 0 |
| La$^-$ | 20.4(2.1) | $4.9 \times 10^4$ | 3103.7 | 96.593 | 1.2 | 3 [a] | 7/2 |

[a] One of the repumping lasers needs to repump three hyperfine levels of the ground state (1487, 1484 and 1455 MHz) according to Ref.8.

**Supplemental Table II.** Measured and optimized binding energies of observed transitions in the present work.

| Labels | Levels (Th$^-\to$ Th) [a] | Measured binding energy (cm$^{-1}$) | Optimized binding energy (cm$^{-1}$) [b] |
|---|---|---|---|
| *T1-1* | $^2S^o_{1/2} \to {}^3F^e_2$ | 782(23) | 783.1(11) |
| *1* | $^4F^o_{7/2} \to {}^3F^e_2$ | 1234(21) | 1235(12) |
| *2* | $^4F^e_{9/2} \to {}^3F^e_2$ | 1738(20) | 1734.5(59) |
| *3* | $^4F^e_{7/2} \to {}^3F^e_2$ | 2007(17) | 2008.4(50) |
| *4* | $^4G^o_{5/2} \to {}^3F^e_2$ | 3054(12) | 3054(12) |
| *5* | $^4F^e_{5/2} \to {}^3F^e_2$ | 3241(11) | 3244(6) |
| *T1-2* | $^2S^o_{1/2} \to {}^3P^e_0$ | 3334(11) | 3341.2(11) |
| *6* | $^4F^o_{7/2} \to {}^3F^e_3$ | 4105(14) | 4104(12) |
| *7* | $^4F^e_{9/2} \to {}^3F^e_3$ | 4596(12) | 4603.6(59) |
| *T1-3* | $^2S^o_{1/2} \to {}^3P^e_1$ | 4643.4(63) | 4648.6(11) |
| *8* | $^4F^e_{7/2} \to {}^3F^e_3$ | 4878.8(61) | 4877.5(50) |
| *9* | $^4F^e_{3/2} \to {}^3F^e_2$ | 4902.6(61) | 4901.35(48) |
| *10* | $^4F^e_{9/2} \to J = 2$ | 5424.7(71) | 5422.5(59) |

[a] The electronic configuration of Th$^-$ is $6d^37s^2$ for the even parity, and $6d^27s^27p$ for the odd parity.
[b] The optimized binding energy values are deduced according to the assignments, the measured binding energy of each transition and the energy levels of the neutral Th using a global optimization.